\documentclass[a4paper,11pt]{article}
\usepackage{pos}
\usepackage{braket}
\usepackage[utf8]{inputenc}
\usepackage{bbm}
\usepackage{amsmath}
\usepackage{graphicx}
\usepackage{wrapfig}

\usepackage{hyperref}
\hypersetup{
    pdftitle={},
    colorlinks=true,
    linkcolor=blue,
    citecolor=blue,
    filecolor=blue,
    urlcolor=blue
}

\usepackage{tikz,tikz-feynman}
\tikzfeynmanset{compat=1.1.0, warn luatex=false}
\usetikzlibrary{arrows, arrows.meta}

\newcommand{\seq}[0]{\mathrm{seq}}

\newcommand{\vvTFF}[1]{\mathcal{F}_{#1 \rightarrow \gamma^* \gamma^*}}
\newcommand{\vrTFF}[1]{\mathcal{F}_{#1 \rightarrow \gamma^* \gamma}}
\newcommand{\rvTFF}[1]{\mathcal{F}_{#1 \rightarrow \gamma \gamma^*}}

\newcommand{\amu}[1]{a_\mu^{#1\text{-pole}}}

\newcommand{\fig}{Figure~}
\newcommand{\eq}{Eq.~}
\newcommand{\eqs}{Eqs.~}

\DeclareMathOperator{\diag}{diag}

\definecolor{emerald}{rgb}{0.31, 0.78, 0.47}

\definecolor{darkamber}{rgb}{1.0, 0.65, 0.0}

\title{Pseudoscalar-pole contributions to the muon \texorpdfstring{$g-2$}{g-2} at the physical point}

\author*[a]{S.~Burri}
\author*[a]{G.~Kanwar}
\author[b,c]{C.~Alexandrou}
\author[c]{S.~Bacchio}
\author[d]{G.~Bergner}
\author[c]{J.~Finkenrath}
\author[a]{A.~Gasbarro}
\author[b,c]{K.~Hadjiyiannakou}
\author[e]{K.~Jansen}
\author[f]{B.~Kostrzewa}
\author[c]{G.~Koutsou}
\author[g]{K.~Ottnad}
\author[h,i]{M.~Petschlies}
\author[c]{F.~Pittler}
\author[h,i]{F.~Steffens}
\author[h,i]{C.~Urbach}
\author[a]{U.~Wenger}

\affiliation[a]{Albert Einstein Center for Fundamental Physics,
  Institute for Theoretical Physics, University of Bern,
  Sidlerstrasse 5,
  CH--3012 Bern, Switzerland}

\affiliation[b]{Department of Physics, University of Cyprus, 20537~Nicosia, Cyprus}

\affiliation[c]{Computation-based Science and Technology Research Center, The Cyprus Institute, 20~Konstantinou Kavafi Street, 2121~Nicosia, Cyprus}

\affiliation[d]{University of Jena, Institute for Theoretical Physics, Max-Wien-Platz 1, D-07743~Jena, Germany}

\affiliation[e]{Deutsches Elektronen-Synchrotron DESY, Platanenallee
  6, 15738 Zeuthen, Germany}

\affiliation[f]{High Performance Computing and Analytics Lab,
  Rheinische Friedrich-Wilhelms-Universit{\"a}t Bonn,
  Friedrich-Hirzebruch-Allee~8,
  D-53115 Bonn, Germany}

\affiliation[g]{PRISMA$^+$ Cluster of Excellence and Institut f{\"u}r Kernphysik,
Johannes Gutenberg-Universit{\"a}t Mainz, Johann-Joachim-Becher-Weg
45, D-55128 Mainz, Germany}

\affiliation[h]{HISKP (Theory), Rheinische
  Friedrich-Wilhelms-Universit{\"a}t Bonn,
  Nussallee~14-16,
  D-53115~Bonn, Germany}

\affiliation[i]{Bethe Center for Theoretical Physics, Rheinische
  Friedrich-Wilhelms-Universit{\"a}t Bonn, Wegelerstraße 10, D-53115 Bonn, Germany}

\emailAdd{burri@itp.unibe.ch}
\emailAdd{kanwar@itp.unibe.ch}

\abstract{Pseudoscalar-pole diagrams are an important component of estimates of the
hadronic light-by-light (HLbL) contribution to the muon $g-2$.
We report on our computation of the transition form factors $\vvTFF{P}$ for the neutral pseudoscalar mesons $P=\pi^0$ and $\eta$. The calculation is performed using twisted-mass lattice QCD with physical quark masses.
On the lattice, we have access to a broad range of (space-like) photon four-momenta and therefore produce form factor data complementary to the experimentally accessible single-virtual direction, which directly leads to an estimate of the pion- and $\eta$-pole components of the muon $g-2$.
For the pion, our result for the $g-2$ contribution in the continuum is comparable with previous lattice and data-driven determinations, with combined relative uncertainties below $10\%$.
For the $\eta$ meson, we report on a preliminary determination from a single lattice spacing.
}

\FullConference{%
  The 39th International Symposium on Lattice Field Theory (Lattice2022),\\
  8-13 August, 2022 \\
  Bonn, Germany 
}

\begin{document}
\maketitle
	
\section{Introduction}
Here we report on the progress of the calculation originally presented in Ref.~\cite{Burri:2021cxr}.
We aim to compute the pseudoscalar transition form
factors $\vvTFF{P}$ from twisted-mass lattice QCD for the three
pseudoscalar states $P=\pi^0, \eta$ and $\eta'$ in order to determine
the corresponding pseudoscalar-pole contributions to 
the hadronic light-by-light (HLbL) scattering in the anomalous
magnetic moment of the muon, $a_\mu = (g_\mu-2)/2$. Presented here are the recent developments of our calculation for the states $P=\pi^0$ and $\eta$.
We employ twisted-mass lattice QCD at
maximal twist, so that we profit from automatic ${\cal O}(a)$ improvement of observables~\cite{Frezzotti:2003ni, Frezzotti:2004wz}. The
pion-pole calculation is performed using three $N_f=2+1+1$ ensembles with varying
lattice spacings, while the preliminary $\eta$-pole calculation uses
the ensemble at the coarsest lattice spacing only.
The production of these ensembles by the Extended Twisted Mass Collaboration (ETMC) is described in Refs.~\cite{Alexandrou:2018egz,ExtendedTwistedMass:2021gbo,ExtendedTwistedMass:2021qui}.
The quark masses for all three ensembles are tuned such that the charged-pion mass is fixed to its physical value and the $s$- and $c$-quark masses approximate their
physical values. A summary of the properties of the three ensembles is presented in Table~\ref{tab:ensembles}.

\begin{table}[b]
  \centering
  \resizebox{\textwidth}{!}{ 
  \begin{tabular}{|l | c | c | c | c | c | c | c |}
    \hline
    ensemble & $L^3 \cdot T / a^4$ & $m_\pi$ [MeV] & $a$ [fm] & $L$ [fm] & $m_\pi \cdot L$ & $Z_V$ & $Z_A$\\ \hline
    cB072.64 & $64^3 \cdot 128$ & 140.2(2) & 0.07961(13) & 5.09 & 3.62 & 0.706378(16) & 0.74284(23)\\
    cC060.80 & $80^3 \cdot 160$ & 136.7(2) & 0.06821(12) & 5.46 & 3.78 & 0.725405(13) & 0.75841(16)\\
    cD054.96 & $96^3 \cdot 192$ & 140.8(2) & 0.05692(10) & 5.46 & 3.90 & 0.744105(11) & 0.77394(10)\\
    \hline
  \end{tabular}
  } %
  \caption{Description of ETMC ensembles used for the analysis presented in
    these proceedings, including the lattice geometry, pion mass $m_\pi$, lattice spacing $a$, lattice size $L$, and the renormalization constants $Z_V$ and $Z_A$ for the vector and axial currents~\cite{Alexandrou:2018egz,ExtendedTwistedMass:2021gbo,ExtendedTwistedMass:2021qui,Alexandrou:2022amy}.
    \label{tab:ensembles}
  }
\end{table}

\begin{figure}[b]
    \centering
\tikzfeynmanset{
  tff/.style={
    /tikz/shape=circle,
    /tikz/draw=black,
    /tikz/pattern=north east lines
  }
}

\tikzfeynmanset{
  blob/.style={
    /tikz/shape=circle,
    /tikz/minimum size=1cm,
    /tikz/draw=black,
    /tikz/pattern=north east lines
  }
}

\begin{tikzpicture}[baseline=(current bounding box.center), scale=1.4]
\begin{feynman}[medium]

\vertex (i) at (0.5,0.5);
\vertex (a) at (1.25,1);
\vertex (b) at (2,1);
\vertex (c) at (2.75,1);
\vertex (o) at (3.5,0.5);
\vertex [tff] (x) at (1.75,2) {};
\vertex [tff] (z) at (2.375,1.5) {};
\vertex (y) at (2,3);

\diagram* {
(i) -- [fermion] (a) -- [fermion] (b) -- [fermion] (c) -- [fermion] (o),
(a) -- [boson] (x) -- [boson] (y),
(b) -- [boson] (z) -- [boson] (c),
(x) -- [scalar, edge label=\(P\)] (z)
};
\end{feynman}
\node at (2,0.5) {+ crossed};
\end{tikzpicture}
\qquad
\begin{tikzpicture}[baseline=(current bounding box.center), scale=1.4]
\begin{feynman}[medium]

\vertex (i) at (0.5,0.5);
\vertex (a) at (1.25,1);
\vertex (b) at (2,1);
\vertex (c) at (2.75,1);
\vertex (o) at (3.5,0.5);
\vertex [tff] (x) at (1.75,2) {};
\vertex [tff] (z) at (2.375,1.5) {};
\vertex (y) at (2,3);

\diagram* {
(i) -- [fermion] (a) -- [fermion] (b) -- [fermion] (c) -- [fermion] (o),
(b) -- [boson, bend left] (x) -- [boson] (y),
(a) -- [boson, bend left] (z) -- [boson] (c),
(x) -- [scalar, edge label=\(P\)] (z)
};
\end{feynman}
\node at (2,0.5) {\phantom{+ crossed}};
\end{tikzpicture}     \caption{The pseudoscalar-pole diagrams contributing to the leading order HLbL scattering in the muon anomalous magnetic moment. Striped circles indicate the nonperturbative $P \rightarrow \gamma^* \gamma^*$ transition form factors required to evaluate these contributions.}
    \label{fig:HLbL pole dominance}
\end{figure}
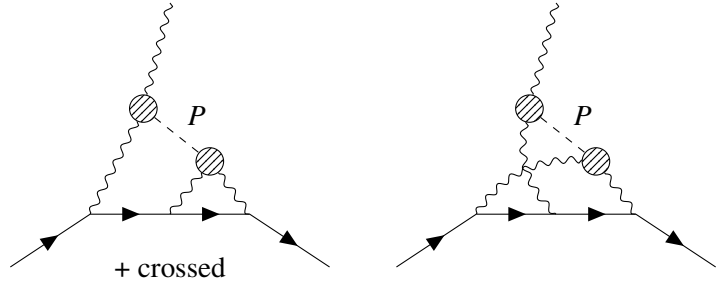

Under the assumption of pole dominance, the leading contributions to the hadronic light-by-light scattering come from exchanges of a neutral pseudoscalar meson $P \in \{ \pi^0, \eta, \eta' \}$. The corresponding contributions to the muon anomalous magnetic moment, $a_\mu^{P\textrm{-pole}}$, are given by the diagrams shown in \fig\ref{fig:HLbL pole dominance}. The nonperturbative information is encapsulated in the
transition form factors $\vvTFF{P}$ of the pseudoscalar meson $P$ to two virtual photons.
The pole contributions are given by a three-dimensional
integral derived in Ref.~\cite{Knecht:2001qf},
\begin{align}
a_\mu^{P\textrm{-pole}} = \left( \frac{\alpha}{\pi} \right)^3 \int_0^\infty & dQ_1\int_0^\infty dQ_2 \int_{-1}^{1} d\tilde\tau \nonumber \\ &\bigg[ w_1(Q_1, Q_2, \tilde\tau)  \vvTFF{P}(-Q_1^2, -Q_3^2) \vrTFF{P}(-Q_2^2, 0) \nonumber \\
	&+w_2(Q_1, Q_2, \tilde\tau)  \vvTFF{P} (-Q_1^2, -Q_2^2) \vrTFF{P} (-Q_3^2, 0) \bigg],
   \label{eq:3drepresentation}
\end{align}
where $Q_3^2 = Q_1^2 + Q_2^2 + 2 \tilde{\tau} Q_1 Q_2$ and the weight functions $w_1$ and $w_2$ are analytically known functions of the kinematics and muon and pseudoscalar masses.
Each term in the integrand in \eq\eqref{eq:3drepresentation} involves two transition form factors (TFFs) at space-like momenta, one at single-virtual kinematics and the other at double-virtual kinematics.
First lattice results for the pion-pole contribution were obtained in Refs.~\cite{Gerardin:2016cqj,Gerardin:2019vio}, while for the $\eta$-pole contribution preliminary results have so far only been presented using
a lattice spacing of $a=0.1315\,\mathrm{fm}$ and lattice size $L = 4.21\,\mathrm{fm}$~\cite{Verplanke2022}.
The pion analysis first presented in Ref.~\cite{Burri:2021cxr} and extended in these proceedings complements prior work
by working directly at the physical point and using a different discretization.
For the $\eta$ meson we provide a result at the physical point with a single lattice spacing of $a=0.0796\,\mathrm{fm}$ with lattice size $L = 5.09\,\mathrm{fm}$; an extension of this result has also now been reported in Ref.~\cite{Alexandrou:2022qyf}.

\section{The transition form factors on the lattice}
Following Refs.~\cite{Gerardin:2016cqj,Burri:2021cxr}, 
the transition form factor in continuum Minkowski space-time is defined via the matrix element of two electromagnetic currents $j_\mu$ and $j_\nu$
between the vacuum and the pseudoscalar state $P$,
\begin{align} \label{eq:tff-from-M}
M_{\mu \nu}(p, q_1) &= i \int d^4x \, e^{i q_1 x} \left<0\left|T\{j_\mu(x) j_\nu(0) \}\right|P(p)\right> \nonumber \\
	&= \varepsilon_{\mu \nu \alpha \beta} q_1^\alpha q_2^\beta
   \vvTFF{P}(q_1^2, q_2^2)\, .
\end{align}
Here $T\{ \cdot \}$ indicates time-ordering, $p$ is the pseudoscalar four-momentum, $q_1$ is the four-momentum of $j_\mu$, and the four-momentum $q_2 = p - q_1$ of $j_\nu$ is enforced by momentum conservation.
For photon virtualities below the threshold for hadron production, the
transition form factor can be analytically continued to Euclidean
space~\cite{Gerardin:2016cqj} and is thus accessible
on the lattice.
In Euclidean space-time, the matrix element can be recovered via
\begin{equation}
M_{\mu \nu}(p, q_1) = i^{n_0} M_{\mu \nu}^E(p, q_1), \quad M_{\mu \nu}^E = - \int_{-\infty}^{\infty} d\tau \, e^{\omega_1 \tau} \tilde{A}_{\mu \nu}(\tau),
\label{eq:Atilde integration}
\end{equation}
where $n_0$ denotes the number of temporal indices in $M_{\mu \nu}$,
$\omega_1$ is the temporal component of the momentum $q_1 = (\omega_1, \vec{q}_1)$, and
$\tilde{A}_{\mu\nu}$ is the Euclidean matrix element
\begin{equation}
\tilde{A}_{\mu\nu}(\tau) \equiv \int d^3\vec{x} e^{-i \vec{q}_1 \cdot \vec{x}} \braket{0 | T\{ j_\mu(\tau,\vec{x}) j_\nu(0) \} | P(\vec{p})} .
\label{eq:Atilde_timeordering}
\end{equation}
On the lattice this function is accessed from the three-point
function 
\begin{equation}
    C_{\mu\nu}(\tau,t_P) \equiv \int d^3\vec{x} \, d^3\vec{y} \, e^{-i \vec{q}_1 \cdot \vec{x}} e^{i \vec{p} \cdot \vec{y}} \braket{0 | T\{j_\mu(\tau,\vec x) j_\nu(0) \mathcal{O}_P^\dag(-t_P, \vec{y})\} | 0}
\label{eq:amplitude}
\end{equation}
via
\begin{equation}
    \tilde{A}_{\mu\nu}(\tau) = \lim_{t_{P} \rightarrow \infty} \frac{2 E_{P}}{Z_{P}} e^{E_{P} t_{P}} C_{\mu\nu}(\tau, t_{P}),
\label{eq:Atilde}
\end{equation}
where $t_P$ is the pseudoscalar insertion time and $Z_P = \langle 0 | \mathcal{O}_P(\vec{0},0) | P(p) \rangle$. 
In this work, the pseudoscalar meson energy $E_P$ and overlap factor $Z_P$ are determined through a separate analysis of two-point functions of the operator $\mathcal{O}_P$.

For the pion, we use the creation operator $\mathcal{O}_{\pi^0}^\dag =
i \bar{\psi} \lambda_3 \gamma_5 \psi$, while for
the $\eta$ meson, we use the creation operator $\mathcal{O}_\eta^\dag = i \bar{\psi} \lambda_8 \gamma_5 \psi$,
where $\lambda_3 = \diag(1, -1, 0)$ and $\lambda_8 = \diag(1, 1, -2) / \sqrt{3}$
are Gell-Mann matrices describing the $\operatorname{SU}(3)$ flavor structure.
The creation operator $\mathcal{O}_\eta^\dag$ has overlap with the
physical $\eta$-meson state, meaning the correct $\eta$-meson
amplitude is projected at large time separations between the operator
and currents, independent of $\eta$-$\eta'$ meson mixing.
The electromagnetic currents are defined for this $N_f = 2+1+1$ calculation as
\begin{equation}
  j_\mu = \frac{2}{3} \bar\psi_u \,\gamma_\mu \, \psi_u - \frac{1}{3} \bar\psi_d \, \gamma_\mu \, \psi_d - \frac{1}{3} \bar\psi_s \, \gamma_\mu \, \psi_s + \frac{2}{3} \bar\psi_c \, \gamma_\mu \, \psi_c.
\end{equation}
The component of the electromagnetic current involving the light quarks can be decomposed into terms with definite isospin, i.e.,
\begin{equation}
\begin{aligned}
  j_\mu^l \equiv \bar\psi_l \,\gamma_\mu \tilde{Q} \, \psi_l &= \frac{1}{6}j^{0,0}_\mu + \frac{1}{2}j^{1,0}_\mu, \\
  j^{0,0}_\mu &\equiv \bar\psi_l\,\gamma_\mu\,\mathbbm{1}\,\psi_l \\
  j^{1,0}_\mu &\equiv \bar\psi_l\,\gamma_\mu\,\sigma_3\,\psi_l,
\end{aligned}
\end{equation}
where $\psi_l = ( \begin{matrix} \psi_u & \psi_d \end{matrix} )$ indicates the light-quark doublet and $\tilde{Q} = \mathrm{diag}(+2/3,\,-1/3)$ is the relevant charge matrix. The current $j^{0,0}_\mu$ has isospin $I=0$ and $j^{1,0}_\mu$ has isospin $I = 0$, $I_z = 0$.
This decomposition allows us to consider only the corresponding isospin preserving parts of the amplitude.
These currents also need to be further renormalized by $Z_V$, due to the
use of local currents instead of conserved (point-split) currents in
the calculation. The renormalization factors for the ensembles used here have been precisely
determined in Ref.~\cite{Alexandrou:2022amy}, as detailed in Table~\ref{tab:ensembles}.

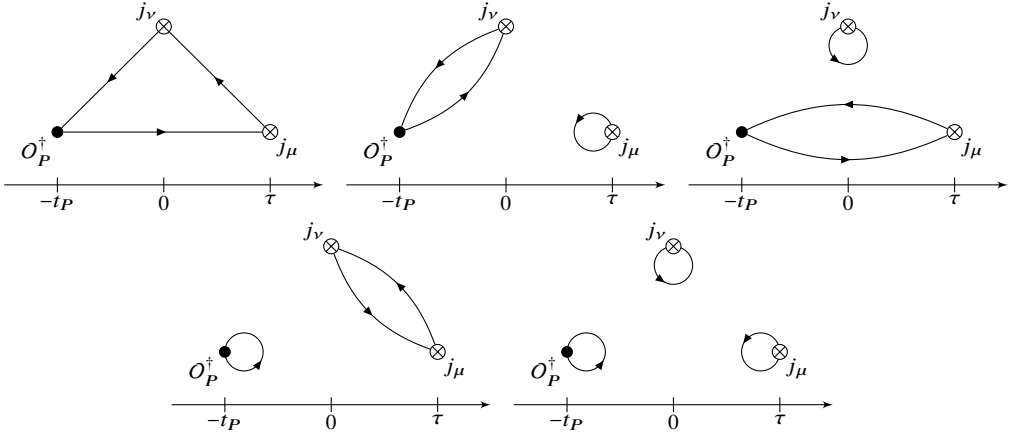
\begin{figure}
    \centering
\newcommand{\tikzTickSize}[0]{2pt}
\newcommand{\tikzJmuX}[0]{1.0}
\newcommand{\tikzJmuY}[0]{1.0}
\newcommand{\tikzJnuX}[0]{2.0}
\newcommand{\tikzAxisY}[0]{-0.5}
\newcommand{\tikzAxisStart}[0]{-0.5}
\newcommand{\tikzAxisEnd}[0]{2.5}
\newcommand{\tikzPX}[0]{0.0}
\newcommand{\tikzPloopX}[0]{0.5}
\newcommand{\tikzTitleY}[0]{1.6}
\newcommand{\tikzFermBend}[0]{25}
\newcommand{\tikzTitleSize}[0]{\footnotesize}
\newcommand{\tikzAxLabelSize}[0]{\scriptsize}
\newcommand{\tikzLabelSize}[0]{\scriptsize}

\tikzfeynmanset{
  cross/.style={/tikz/path picture={ 
    \draw[black]
    (path picture bounding box.south east) -- (path picture bounding box.north west) (path picture bounding box.south west) -- (path picture bounding box.north east);
  }},
  current/.style={
    /tikz/circle,
    /tikz/fill=white,
    /tikz/draw=black,
    /tikz/inner sep=0.7mm,
    /tikzfeynman/cross
  },
  ps/.style={
    /tikz/circle,
    /tikz/draw=black,
    /tikz/fill,
    /tikz/inner sep=0.5mm
  },
  tiny/.style={
    /tikzfeynman/arrow size=0.8pt
  }
}

\newcommand{\wickDiagram}[2]{
\begin{tikzpicture}[scale=1.4]

#2

\begin{feynman}[tiny]

\vertex [ps, label={[inner sep=0]south west:\tikzLabelSize \( \mathcal{O}^\dag_{P} \)}] (p) at (\tikzPX,0) {};
\vertex [current, label={[inner sep=0]north west:\tikzLabelSize \( j_\nu \)}] (mu) at (\tikzJmuX,\tikzJmuY) {};
\vertex [current, label={[inner sep=0]south east:\tikzLabelSize \( j_\mu \)}] (nu) at (\tikzJnuX,0) {};
\vertex (ploop) at (\tikzPloopX,0);

\diagram* {
#1
};

\end{feynman}

\draw[-latex'] (\tikzAxisStart,\tikzAxisY) -- (\tikzAxisEnd,\tikzAxisY);
\draw ($(\tikzPX,\tikzAxisY) + (0,-\tikzTickSize)$) -- ($(\tikzPX,\tikzAxisY) + (0,\tikzTickSize)$) node [below = 1mm] {\tikzAxLabelSize $-t_{P}$};
\draw ($(\tikzJmuX,\tikzAxisY) + (0,-\tikzTickSize)$) -- ($(\tikzJmuX,\tikzAxisY) + (0,\tikzTickSize)$) node [below = 1mm] {\tikzAxLabelSize $0$};
\draw ($(\tikzJnuX,\tikzAxisY) + (0,-\tikzTickSize)$) -- ($(\tikzJnuX,\tikzAxisY) + (0,\tikzTickSize)$) node [below = 1mm] {\tikzAxLabelSize $\tau$};

\end{tikzpicture}
}

\newcommand{\drawPloop}[0]{
\draw [/tikzfeynman/tiny,postaction={/tikzfeynman/with arrow=0.4}] (p) arc (180:540:1.8mm);
}
\newcommand{\drawJmuloop}[0]{
\draw [/tikzfeynman/tiny,postaction={/tikzfeynman/with arrow=0.4}] (mu) arc (90:450:1.8mm);
}
\newcommand{\drawJnuloop}[0]{
\draw [/tikzfeynman/tiny,postaction={/tikzfeynman/with arrow=0.4}] (nu) arc (0:360:1.8mm);
}

\wickDiagram{(p) -- [fermion] (nu) -- [fermion] (mu) -- [fermion] (p)}{
}
\wickDiagram{(p) -- [fermion, bend right=\tikzFermBend] (mu) -- [fermion, bend right=\tikzFermBend] (p)}{
\drawJnuloop
}
\wickDiagram{(p) -- [fermion, bend right=\tikzFermBend] (nu) -- [fermion, bend right=\tikzFermBend] (p)}{
\drawJmuloop
}
\wickDiagram{(mu) -- [fermion, bend right=\tikzFermBend] (nu) -- [fermion, bend right=\tikzFermBend] (mu)}{
\drawPloop
}
\wickDiagram{}{
\drawPloop
\drawJmuloop
\drawJnuloop
}
     \caption{Wick contractions contributing to $C_{\mu\nu}(\tau,t_{P})$. Connected (top left), vector-current disconnected (``V-disconnected'', top middle and right), pseudoscalar disconnected (``P-disconnected'', bottom left) and fully disconnected (bottom right). The second connected diagram with quark propagators running in the opposite direction is omitted for brevity. 
    }
    \label{fig:3pt-diagrams}
\end{figure}

The evaluation of the three-point function $C_{\mu \nu}$ involves connected, vector-current disconnected, pseudoscalar
disconnected, and fully disconnected 
Wick contractions, as illustrated from top left to bottom right in \fig\ref{fig:3pt-diagrams}.
In the case of the $\pi^0$, the amplitude can be related in the isospin-symmetric limit to a charged-pion amplitude 
by the isospin rotation
\begin{equation}
\begin{aligned}
  \pi^0 &\rightarrow -i \cdot (\pi^+ + \pi^-) \, , \\
  j^{0,0}_\mu &\rightarrow j^{0,0}_\mu \, , \\
  j^{1,0}_\mu &\rightarrow i \cdot (j^{1,+}_\mu - j^{1,-}_\mu) \, ,
\end{aligned}
\end{equation}
where $j_\mu^{1,\pm} = \bar\psi_l\,\gamma_\mu\,\sigma_\pm\,\psi_l$ with $\sigma_{\pm} = (\sigma_1 \pm i \sigma_2)/2$.
Note that the currents $j_\mu^{1,\pm}$, when working in the twisted basis, need to be renormalized with $Z_A$ rather than $Z_V$. The values of these renormalization constants are given in Table~\ref{tab:ensembles}.
The isospin rotation simplifies the evaluation of the three-point function by removing the pseudoscalar-disconnected diagram and is thus employed in our analysis of the pion TFF. 
Although isospin symmetry is broken by $O(a^2)$ lattice artifacts in
the twisted-mass Wilson fermion discretization, the effect of this
rotation is removed in the continuum extrapolation. In the case of the
$\eta$- and $\eta'$-meson states, pseudoscalar-disconnected diagrams
do not cancel even in the isospin symmetric limit and indeed play an
important role in capturing the physics of these states.

\begin{figure}
\centering
\includegraphics[width=0.45\textwidth]{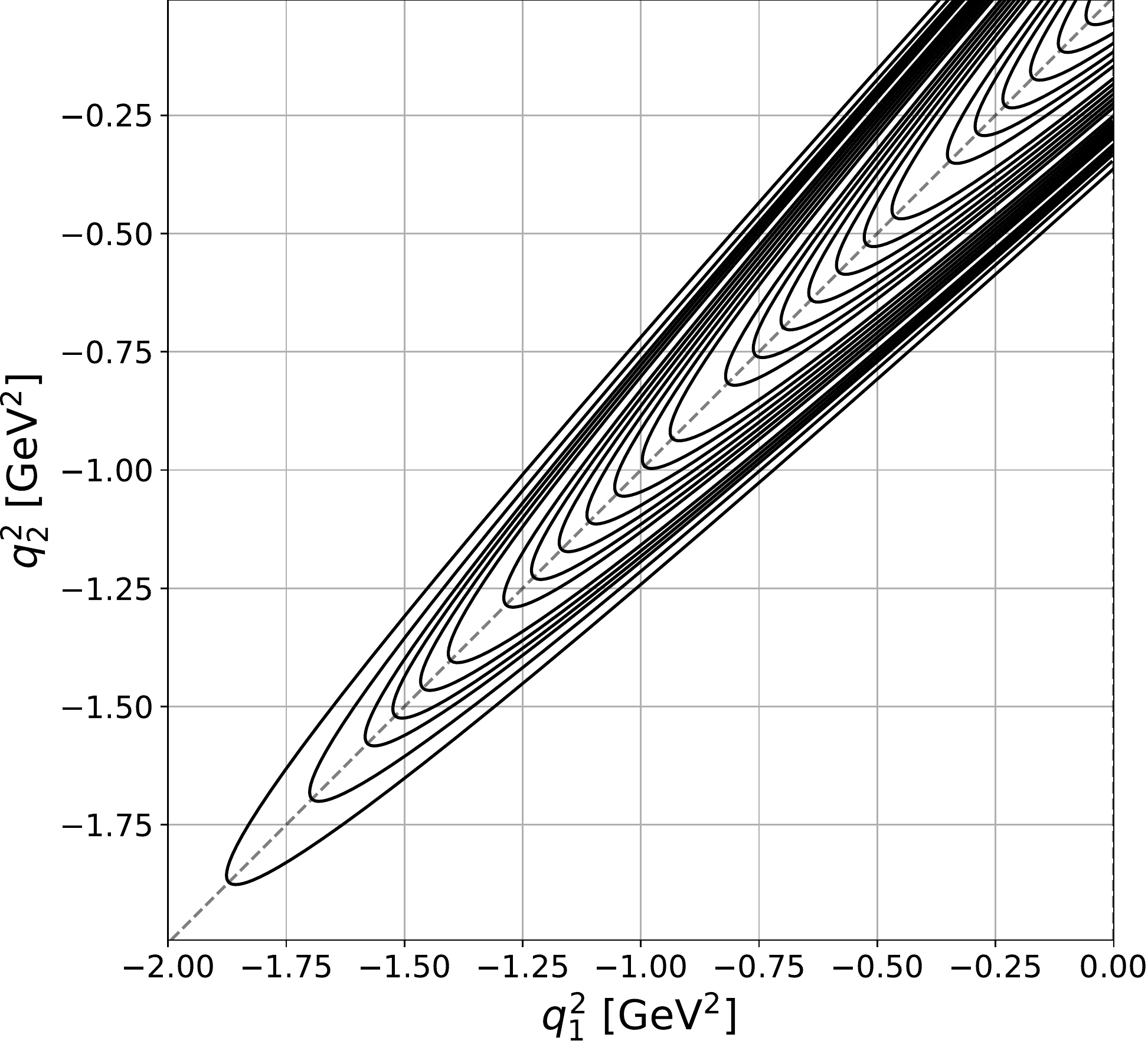}
\includegraphics[width=0.45\textwidth]{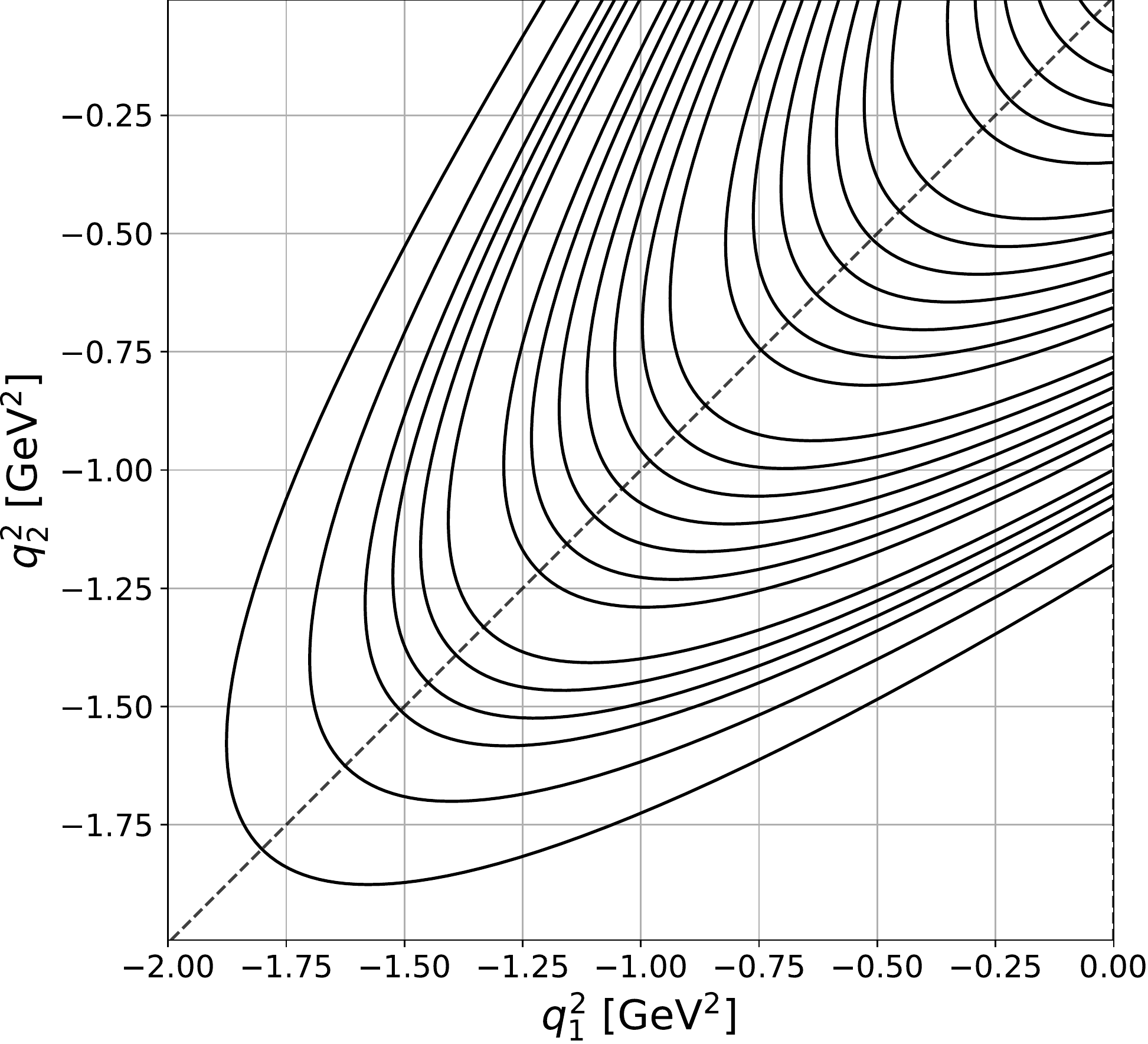}
\caption{Range of photon virtualities for the pion (left) and $\eta$ meson (right) TFFs spanned in our calculation on the ensemble cB072.64. \label{fig:yield_plot_cB_phys}}
\end{figure}

To evaluate the connected diagram, we choose to perform a sequential
inversion through the pseudoscalar operator. This makes it
computationally convenient to project to various choices of momenta
$\vec{q}_1$ for the $j_\mu$ current at the cost of having to restrict
to only a few momenta $\vec{p}$ for the pseudoscalar meson.
The calculation presented here is restricted to pseudoscalar mesons at rest, i.e., $\vec{p} = \vec{0}$. In this kinematic setup, the expressions for the photon virtualities simplify to
\begin{equation}
q_1^2 = \omega_1^2 - \vec{q}_1^2\,, \quad  
        q_2^2 = (m_P - \omega_1)^2 - \vec{q}_1^2  \, .        \label{eq:orbits}
\end{equation}
\noindent
Thus each choice of spatial momentum $\vec q_1$ corresponds to a continuous set of combinations of $q_1$ and $q_2$ which form an orbit in the $(q_1^2,q_2^2)$-plane. The resulting reach in the kinematics of both the pion and $\eta$-meson TFFs on the cB072.64 ensemble is shown in \fig\ref{fig:yield_plot_cB_phys}. The narrowness of the orbits available for the pion TFF is due to the relatively small mass of the pion.
This illustrates the challenge in extracting single-virtual pion
transition form factors $\vrTFF{\pi}(q^2,0) = \rvTFF{\pi}(0,q^2)$ at
significantly negative virtualities $q^2$ on physical point ensembles
if one uses only pions at rest. In future extensions of the analysis
presented here, it would thus be useful to extend the single-virtual
coverage by using moving frames~\cite{Gerardin:2019vio}. For the
$\eta$ meson, due to the higher mass, the problem is less eminent.

The rest frame also simplifies the procedure in \eq\eqref{eq:tff-from-M} of extracting the Lorentz-scalar TFF from the components of the transition amplitude. In particular, one finds in the rest frame that $\tilde A_{\mu \nu}$, and therefore $M_{\mu\nu}$, vanishes when one or more of the indices are temporal. Meanwhile, the spatial components can be averaged to yield the Lorentz scalar $\tilde A(\tau) = i m_P^{-1} \varepsilon_{ijk} (\vec{q}_1^i / |\vec{q}_1|^2) \tilde A_{jk}(\tau)$, which is the appropriate combination to yield the TFF after integration,
\begin{equation} \label{eq:Atilde-avg-integration}
  \vvTFF{P}(q_1^2, q_2^2)|_{\vec{p} = 0} = \int_{-\infty}^{\infty} d\tau e^{\omega_1 \tau} \tilde{A}(\tau).
\end{equation}

\begin{figure}
\centering
\includegraphics[page = 1, width = 0.49\textwidth]{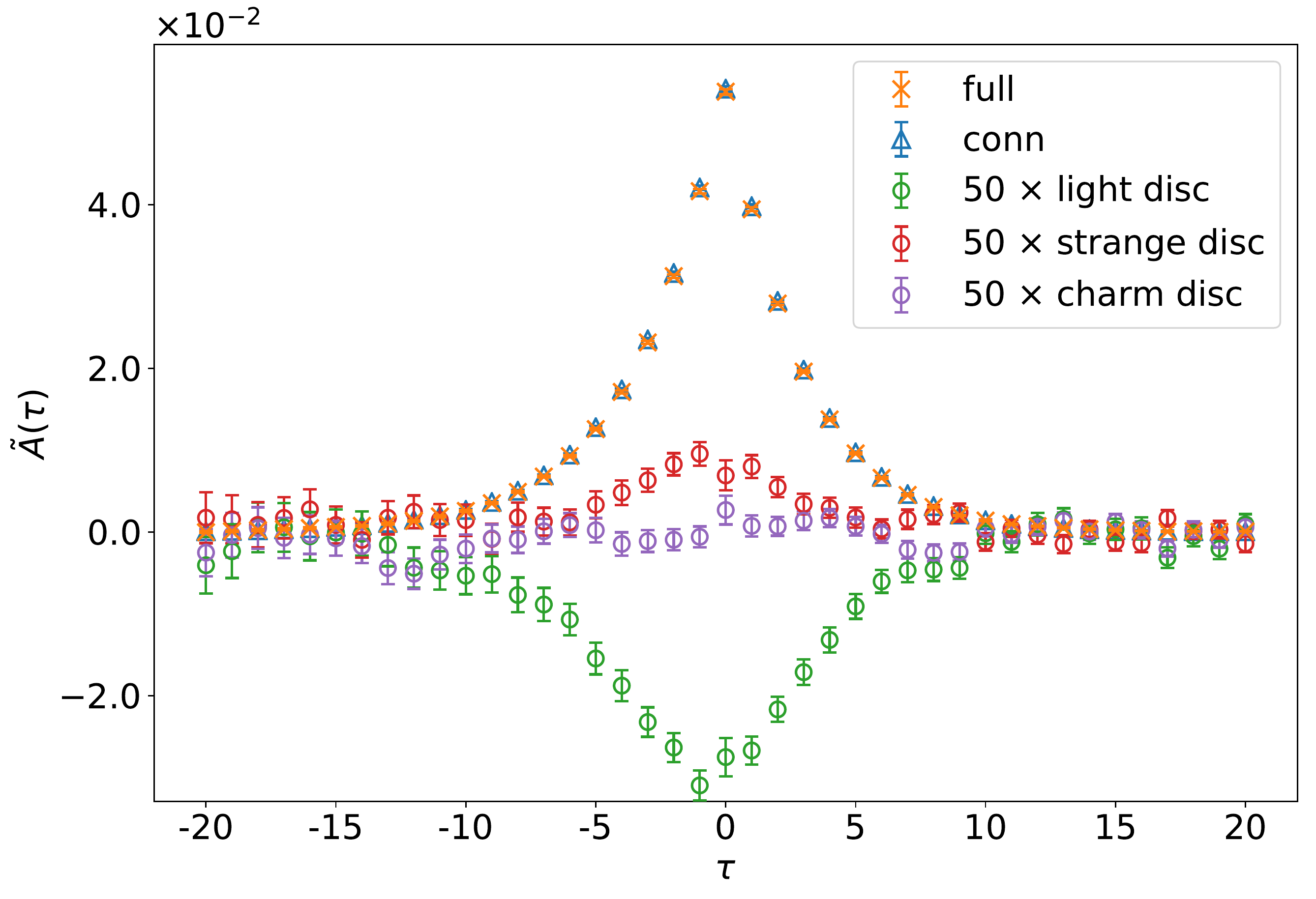}\hfill
\includegraphics[page = 2, width = 0.49\textwidth]{figs/pion/amplitude_tilde_A/plots_divided_wraparound_proceedings_flipped}
\caption{Amplitude $\tilde A(\tau)$ for the pion for momentum orbit $|\vec{q}^2| =
  10\,(2\pi/L)^2$ (left) and $|\vec{q}^2| = 29\,(2\pi/L)^2$ on
  cB072.64. Shown in orange is the full contribution to $\tilde A(\tau)$, in blue the
  connected contribution and in green, red and purple the V-disconnected contributions multiplied by 50.\label{fig:plot_Atilde_pion}
}
\end{figure}

For the pion, examples for the averaged amplitude $\tilde A(\tau)$ are shown in \fig\ref{fig:plot_Atilde_pion}, 
illustrating the full amplitude and separately the connected and the vector-current disconnected contributions with disconnected light-, strange- and charm-current loops for two orbits on the ensemble cB072.64. The vector-current disconnected contributions are multiplied by a factor of 50 to make the comparison to the connected contribution and full amplitude easier. While the disconnected contributions are very small, we still are able to extract significant signals across all momentum orbits for the V-disconnected light and strange contributions. The charm contributions are also determined to sufficient precision to constrain their contribution to the amplitude. We find on all three ensembles that the total V-disconnected contribution in the peak region is suppressed with respect to the connected contribution by a factor of at least $\sim 50$. We also find that the statistical error on the V-disconnected contribution is well under control on all three physical point ensembles.

\begin{figure}
\centering
\includegraphics[page = 1, width = 0.49\textwidth]{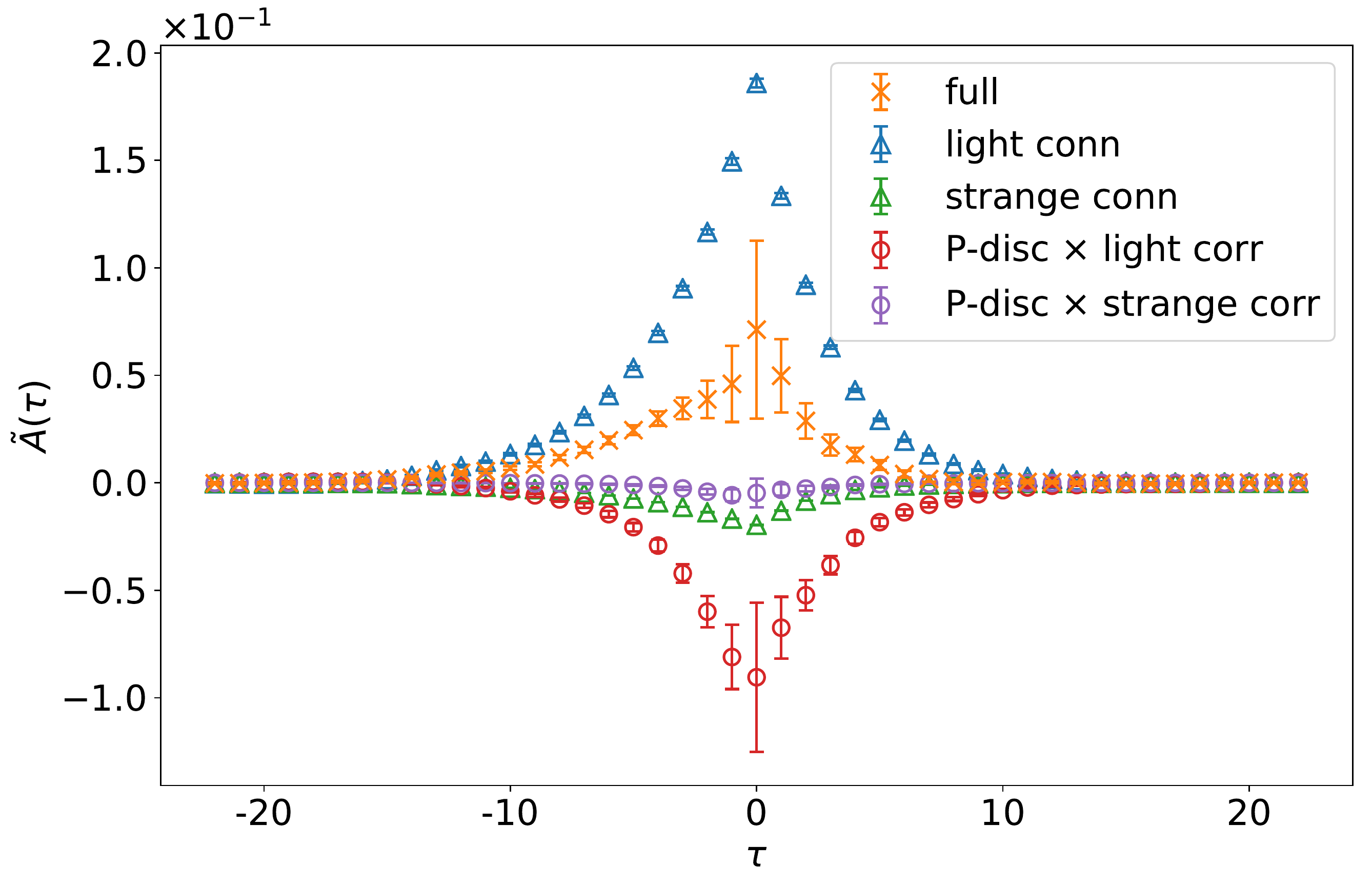}\hfill
\includegraphics[page = 2, width = 0.49\textwidth]{figs/eta/amplitude_tilde_A/plots_cB211_tilde_A_fixed_mass_and_overlap_full_proceedings_flipped}
\caption{Amplitude $\tilde A(\tau)$ for the $\eta$ meson for momentum orbit $|\vec{q}^2| =
  10\,(2\pi/L)^2$ (left) and $|\vec{q}^2| = 29\,(2\pi/L)^2$ on
  cB072.64. Shown is the full value of $\tilde A(\tau)$, as well as the decomposition into the light and strange connected contributions and the light and strange P-disconnected contributions.\label{fig:plot_Atilde_eta}
}
\end{figure}

Sample results for the $\eta$-meson amplitude
$\tilde{A}(\tau)$ as defined in \eq\eqref{eq:Atilde} are displayed in \fig\ref{fig:plot_Atilde_eta} 
where we show the full amplitude and separately 
the different contributions to it considered in our analysis 
for two orbits on the ensemble cB072.64. The examples show that the contributions involving strange quarks in the currents are suppressed at least by a factor of $\sim 10$ compared to the corresponding light contributions. Based on the results in the pion case, we expect the contributions involving charm quarks to be more suppressed still. This contribution along with the vector-current disconnected and fully disconnected diagrams are therefore irrelevant at the presently achievable precision, and we do not include them in the further $\eta$-meson analysis. 

As shown in \eq\eqref{eq:Atilde-avg-integration}, the transition form factors $\vvTFF{P}$ are obtained by integrating $e^{\omega_1 \tau} \tilde A(\tau)$ over the whole temporal axis.
For choices of $\omega_1$ that result in kinematics near the single-virtual axes (cf.~\fig\ref{fig:yield_plot_cB_phys}), the exponential factor $e^{\omega_1 \tau}$ enhances the contribution from one of the tails
exponentially. Meanwhile,
the signal-to-noise ratio deteriorates exponentially at large $|\tau|$, presenting a significant obstacle to the extraction of the TFFs in the single-virtual regime.
Further, the lattice data for times $\tau < -t_P$ have incorrect time ordering of the operators and
do not yield a valid approximation to $\tilde A_{\mu\nu}(\tau)$ in \eq\eqref{eq:Atilde_timeordering}.

To address both of these issues, we extend $\tilde A(\tau)$ by fitting the lattice data with a model function $\tilde A^\textrm{(fit)}(\tau)$ in a range $\tau_\textrm{min} \leq |\tau| < \tau_\textrm{max}$, and replacing the lattice data $\tilde A^\textrm{(latt.)}(\tau)$ by the fit for $|\tau| > \tau_\textrm{cut}$. The integration in \eq\eqref{eq:Atilde-avg-integration} is then replaced by
\begin{equation}
  \vvTFF{P}(q_1^2,q_2^2) = \int_{-\tau_\textrm{cut}}^{\tau_\textrm{cut}} d\tau \, \tilde A^\textrm{(latt.)}(\tau) e^{\omega_1 \tau} + \int_{\tau_\textrm{cut}}^{\infty} d\tau \, \tilde A^\textrm{(fit)}(\tau) e^{\omega_1 \tau} + \int_{-\infty}^{-\tau_\textrm{cut}} d\tau \, \tilde A^\textrm{(fit)}(\tau) e^{\omega_1 \tau}. \label{eq:TFF_fit}
\end{equation}
Note that in principle $\tau_\textrm{cut}$ can be chosen independently
in the two tails in order to keep as much of the original lattice data as
possible. 
For the pion, the contributions to $\vvTFF{\pi}$ from $\tilde
A^{\pi,\textrm{ (fit)}}(\tau)$ are below 2\% for most photon
virtualities, with the exception of some large virtualities
at or close to single-virtual kinematics.
In the case of the $\eta$ meson,
the creation operator insertion happens at Euclidean times $t_{\eta}$ much closer to zero,
as compared to the pion, to control the rapidly deteriorating signal-to-noise ratio for the P-disconnected contribution with increasing $t_{\eta}$ and $\tau$. This requires smaller values of $\tau_{\textrm{cut}}$ to satisfy correct time-ordering of the amplitude,
resulting in larger contributions from $\tilde A^{\eta,\textrm{ (fit)}}(\tau)$ to $\vvTFF{\eta}$
despite the much faster exponential decay of the heavier $\eta$-meson.

\begin{figure}
  \includegraphics[page=2, width = 0.49\textwidth]{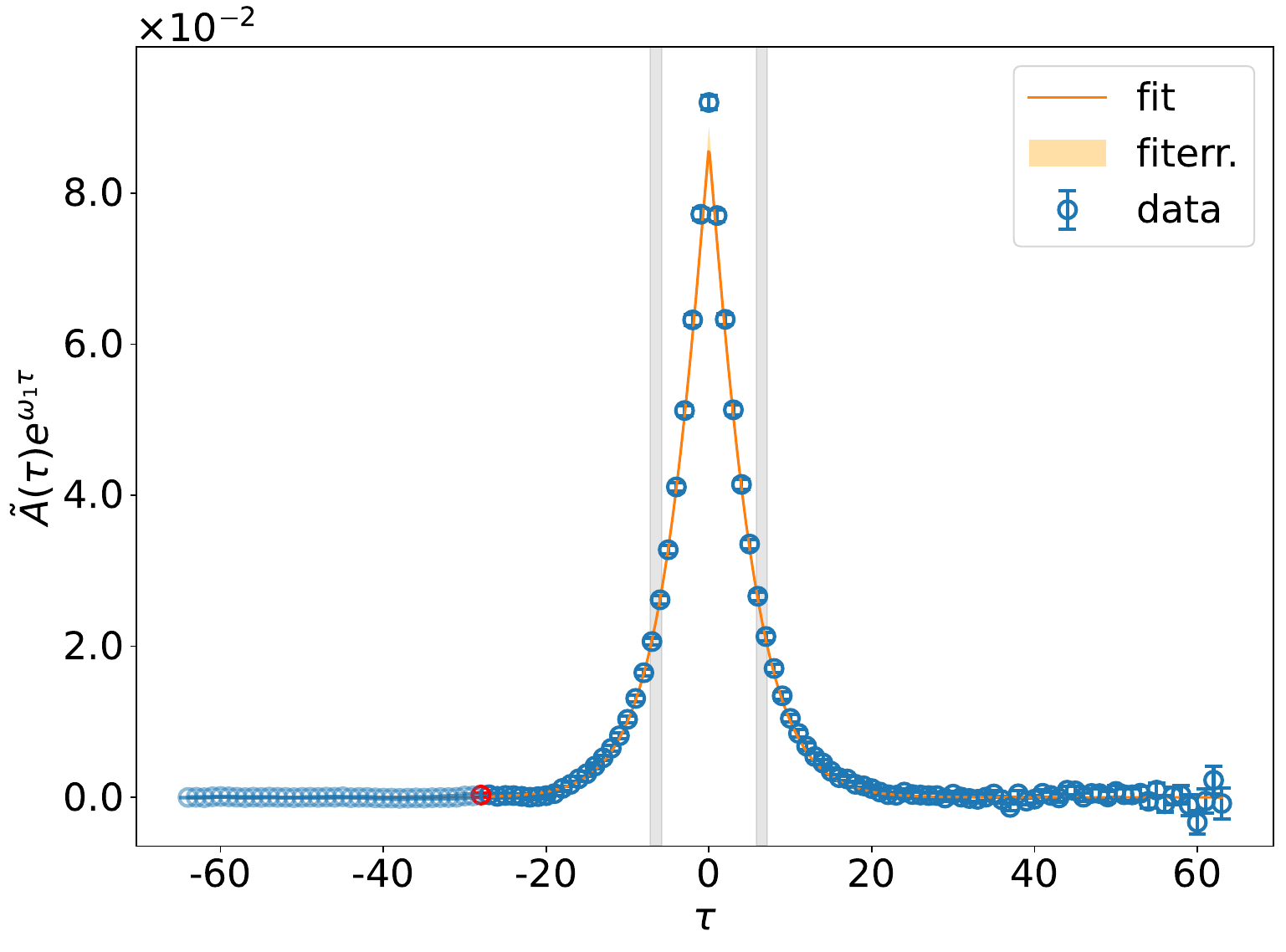} \hfill
  \includegraphics[page=2, width = 0.49\textwidth]{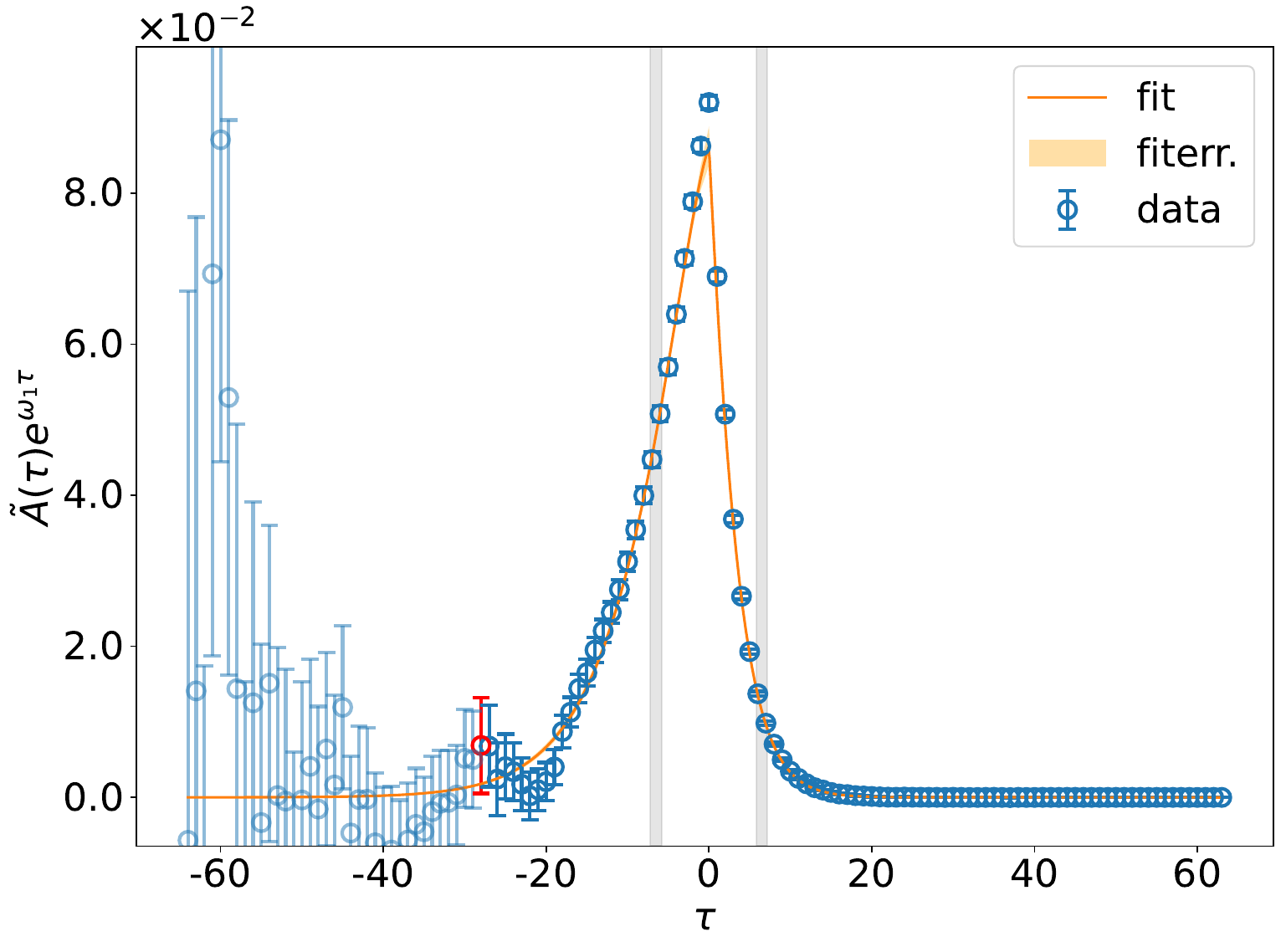}
  \caption{
    Lattice data vs.\ the LMD model fit of the integrand
    $\tilde A(\tau) e^{\omega_1 \tau}$ for the pion on the cB072.64 ensemble
    with the momentum of the current $|\vec{q}_1^2| = 2\,(2\pi/L)^2$.
    The left plot shows the integrand for diagonal kinematics ($q^2_1 = q^2_2$) determined by fixing $a \omega_1 = a m_\pi/2 \approx 0.0283$,
    while the right plot shows the integrand for single-virtual kinematics ($q^2_2 = 0$) determined by fixing  $a \omega_1 = a |\vec{q_1}| \approx 0.1388$. The grey bands show the
    data included in the fit from this choice of $\vec{q}_1$; note, however, that data within these windows are taken across all choices of $\vec{q}_1$ for the fit. The point in red indicates the timeslice where the
    pseudoscalar creation operator is inserted, with the lighter points to the left of
    this insertion having incorrect time ordering, as discussed in the main text. 
    \label{fig:global_LMD_fit_pion} }
\end{figure}

Following the approach of Ref.~\cite{Gerardin:2016cqj}, we consider
both the vector-meson dominance (VMD) model and the lowest-meson
dominance (LMD) model to fit the amplitude. The variation between the two then gives an
estimate of the model dependence of the results.
For the pion we perform global fully correlated fits, i.e., we
simultaneously fit all momentum orbits in the range $\tau_\textrm{min}
\leq |\tau| < \tau_\textrm{max}$ and take into account the
correlations between all fitted data, while for the $\eta$ meson we consider both global uncorrelated and fully correlated fits. An example of this is shown in \fig\ref{fig:global_LMD_fit_pion} 
where we plot the integrand $\tilde A(\tau)e^{\omega_1 \tau}$
resulting from a fully correlated global fit to $\tilde A(\tau)$ using the LMD model in the range $6 \leq |\tau/a| < 8$ with $\chi^2/\textrm{dof}=1.08$ for the pion on the ensemble cB072.64. The fit quality and reduced $\chi^2$ are representative of the remaining fits performed for other analysis choices.
Shown are the integrands for diagonal kinematics $q_1^2=q_2^2$ on the left
and single-virtual kinematics $q_1^2=0$ on the right.
Note that the so-obtained values of the TFFs depend on the choice of
the model, the fit range and the value of $\tau_\textrm{cut}$. These various
choices are all independently carried through the further analysis steps and are finally used to estimate the systematic error of $a_\mu^{P\textrm{-pole}}$.

\section{Results for \texorpdfstring{$a_\mu^{P-\mathrm{pole}}$}{the anomalous pseudoscalar pole contributions} at the physical point}
To extend the form factors to arbitrary photon momenta for the integration in \eq\eqref{eq:3drepresentation}, we parameterize them using a model-independent expansion of the form~\cite{Gerardin:2019vio}
\begin{multline}
P(Q_1^2, Q_2^2) \cdot \mathcal{F}_{P \gamma^* \gamma^*} (-Q_1^2, -Q_2^2) \approx \\
 \sum_{m,n = 0}^N c_{nm} \left(z_1^n - (-1)^{N+n+1} \frac{n}{N+1}
   z_1^{N+1} \right) \left(z_2^m - (-1)^{N+m+1} \frac{m}{N+1}
   z_2^{N+1} \right),
 \label{eq:modified z-expansion}
\end{multline}
termed the ``modified $z$-expansion.''
Above, $P(Q_1^2, Q_2^2) = 1 + (Q_1^2 + Q_2^2)/M_\rho^2$ preconditions the form of the expansion to more easily reproduce the form factor structure, the coefficients $c_{nm} = c_{mn}$ are required to be symmetric by the Bose symmetry of the TFF, and
the $z_k$ are conformal transformations of the four-momenta in the quadrant below the non-analytic cuts at $t_c = 4 m_\pi^2$, as given by
\begin{equation}
  z_k = \frac{\sqrt{t_c + Q_k^2} - \sqrt{t_c - t_0}}{\sqrt{t_c + Q_k^2} + \sqrt{t_c - t_0}}.
\end{equation}
The parameter $t_0$ is chosen to be
\begin{equation}
    t_0 = t_c \left(1 - \sqrt{1 + Q_{\mathrm{max}}^2 / t_c} \right)
\end{equation}
in order to best reproduce the behavior of the TFF for $Q_{1,2}^2 \lesssim Q_{\mathrm{max}}^2$~\cite{Gerardin:2019vio}, with $Q_{\mathrm{max}}^2 = 4.0 \, \mathrm{GeV}^2$ chosen for the present study. Analyticity of the TFF below $t_c$ ensures that the $z$-expansion can fully describe the TFF as $N \rightarrow \infty$.
To fix the coefficients $c_{nm}$ in the expansion, a fit is performed to the lattice data available at the orbits determined by our choices of three-momenta $\vec{q}_1$ (cf.~\fig\ref{fig:yield_plot_cB_phys}).
Because the orbits are in principle continuously determined by the
free choice of $\omega_1$, we follow
Refs.~\cite{Gerardin:2016cqj,Gerardin:2019vio} and fix a finite set of inputs $\vvTFF{P}(-Q_1^2,-Q_2^2)$ by restricting to kinematics satisfying $Q_2^2 / Q_1^2 = \textrm{const.}$ for several choices of this ratio. We further restrict the inputs in the pion case by removing points for which the model contribution to $\vvTFF{\pi}$ from $\tilde A^\textrm{(fit)}(\tau)$ exceeds a given threshold, thereby minimizing the impact of the choice of fit model at this stage of the analysis. In the following analysis, both a threshold of $5\%$ and $10\%$ are considered. We use the smaller choice of $5\%$ for the final continuum-extrapolated result.
\subsection{Pion}
\begin{figure}
\centering
\includegraphics[page=1, width =
1.0\textwidth]{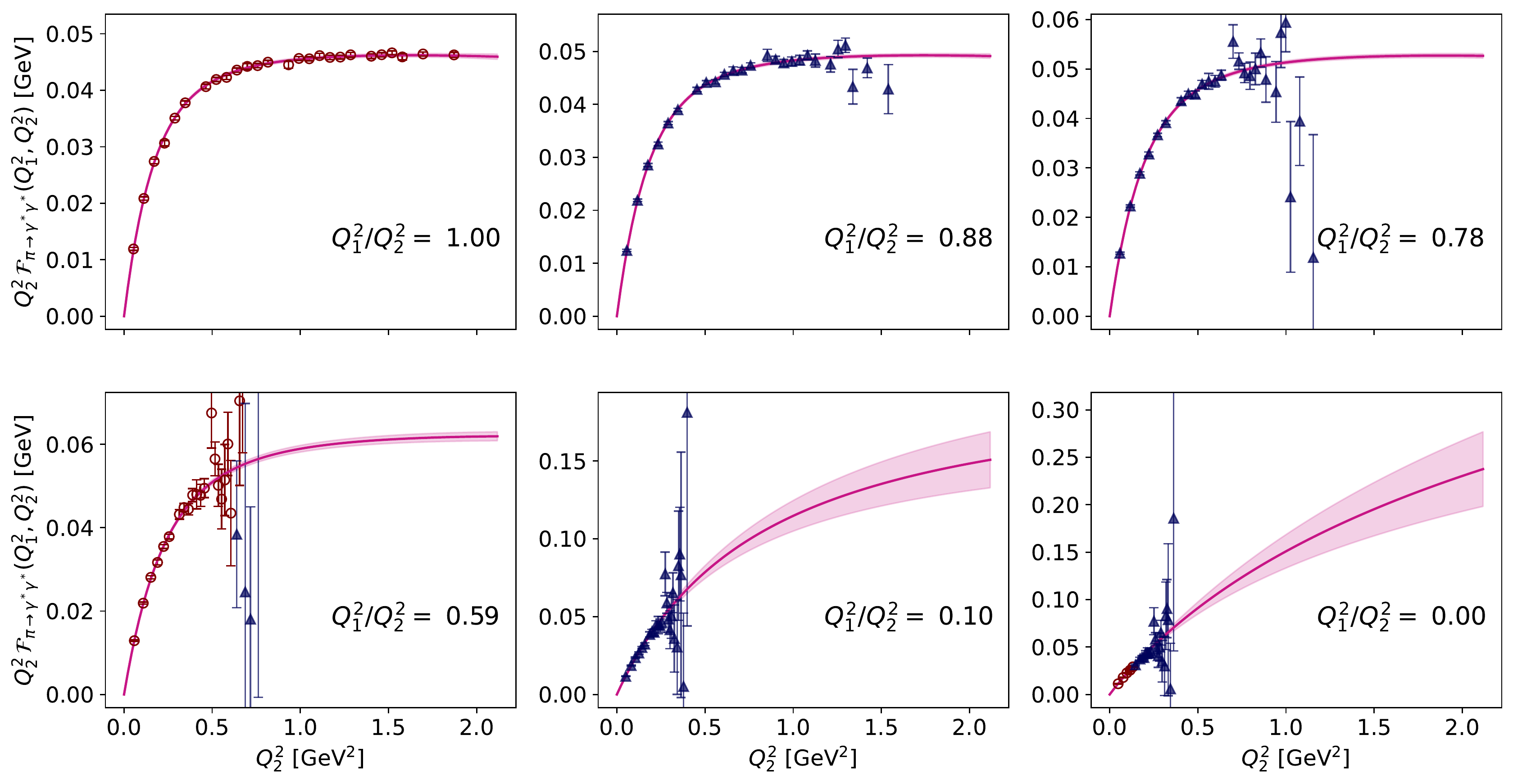}
\caption{Illustration of the pion transition form factor $\vvTFF{\pi}$ and its
  parameterization using the fitted modified $z$-expansion. Only the data
coloured in maroon is included in the fit. %
\label{fig:zExpansion_pion}}
\end{figure}

\fig\ref{fig:zExpansion_pion} 
shows the example of an $N=2$ $z$-expansion fit using TFF data satisfying $Q_2^2/Q_1^2 \in \{1.0, 0.59, 0.0\}$ on the
ensemble cB072.64, where the transition form factors are obtained from a global LMD fit with $\{\tau_\textrm{min}/a, \tau_\textrm{max}/a\} = \{9, 12 \}$ and integration using
$\tau_\textrm{cut}/a$ = 20. Here we use a maximal upper limit of a $10\%$ contribution to $\vvTFF{\pi}$ from $\tilde A^\textrm{(fit)}(\tau)$ as a further cut on the input data for the fit. The points marked with maroon in the figure indicate the final data used in the fit, with other points, including the data for the ratios $Q_2^2/Q_1^2 \in \{0.88, 0.78, 0.10\}$, serving as a crosscheck for the quality of the fit. Both the fit to $\tilde A(\tau)$ and the fit to the modified $z$-expansion are fully correlated, with respective $\chi^2_{z-\mathrm{exp}}/\textrm{dof} = 1.03$ and $\chi^2_{\mathrm{LMD}}/\textrm{dof} = 1.03$.
It is worth noting that the $z$-expansion fit is most strongly constrained by the precise data available for the ratio $Q_2^2/Q_1^2=1.0$. This precision is due to a small dependence on the tails of the integrand $\tilde A(\tau) \exp(\omega_1 \tau)$ for these kinematics (cf.~\fig\ref{fig:global_LMD_fit_pion}).

After extrapolating $\vvTFF{\pi}(-Q_1^2, -Q_2^2)$
 to arbitrary space-like momenta using the results of
the modified $z$-expansion fits in \eq\eqref{eq:modified z-expansion},
the three-dimensional integral representation given in
\eq\eqref{eq:3drepresentation} can be evaluated to calculate $\amu{\pi}$. We do
so for $\mathcal{O}(1000)$ combinations of the choices of fit range and
fit model in the $\tilde A(\tau)$ fit, the choice of
$\tau_\textrm{cut}$, and the choices of samplings of $\vvTFF{\pi}$ in the $(Q_1^2,Q_2^2)$-plane
as inputs to the modified $z$-expansion fit. For the pion,
the fully correlated $z$-expansion fits with $N = 2$
give the best reduced $\chi^2$ values, so we restrict to this choice alone.
Further, for the pion we consider choices of $\tau_\textrm{cut}$
satisfying $\tau_\textrm{cut} \in [1.3, 1.6]\,\mathrm{fm}$ in physical
units across all three ensembles.
We then use a modified version of the Akaike information criterion
(AIC) to perform a weighted average across the analysis choices; see
Ref.~\cite{Borsanyi:2020mff} and references therein. Using this procedure
we obtain separate estimates of the statistical errors and systematic errors
associated with analysis choices described above.
The preliminary values for $\amu{\pi}$ obtained per ensemble are presented in Table~\ref{tab:AIC_results}.
\begin{table}
  \centering
  \begin{tabular}{|l | c | c | c |}
    \hline
    & cB072.64 & cC060.80 & cD054.96 \\ \hline
    $\amu{\pi} \cdot 10^{11}$ & 56.8(2.7)(0.3)[2.7] & 56.5(2.1)(0.5)[2.2] & 54.0(2.1)(0.3)[2.2] \\
    \hline
  \end{tabular}
  \caption{Preliminary results for estimates of $\amu{\pi}$ on the three ensembles used in this analysis. The uncertainties shown are statisical, systematic and total, respectively.
    \label{tab:AIC_results}
  }
\end{table}

\begin{figure}
\centering
\includegraphics[page=2, width =
0.7\textwidth]{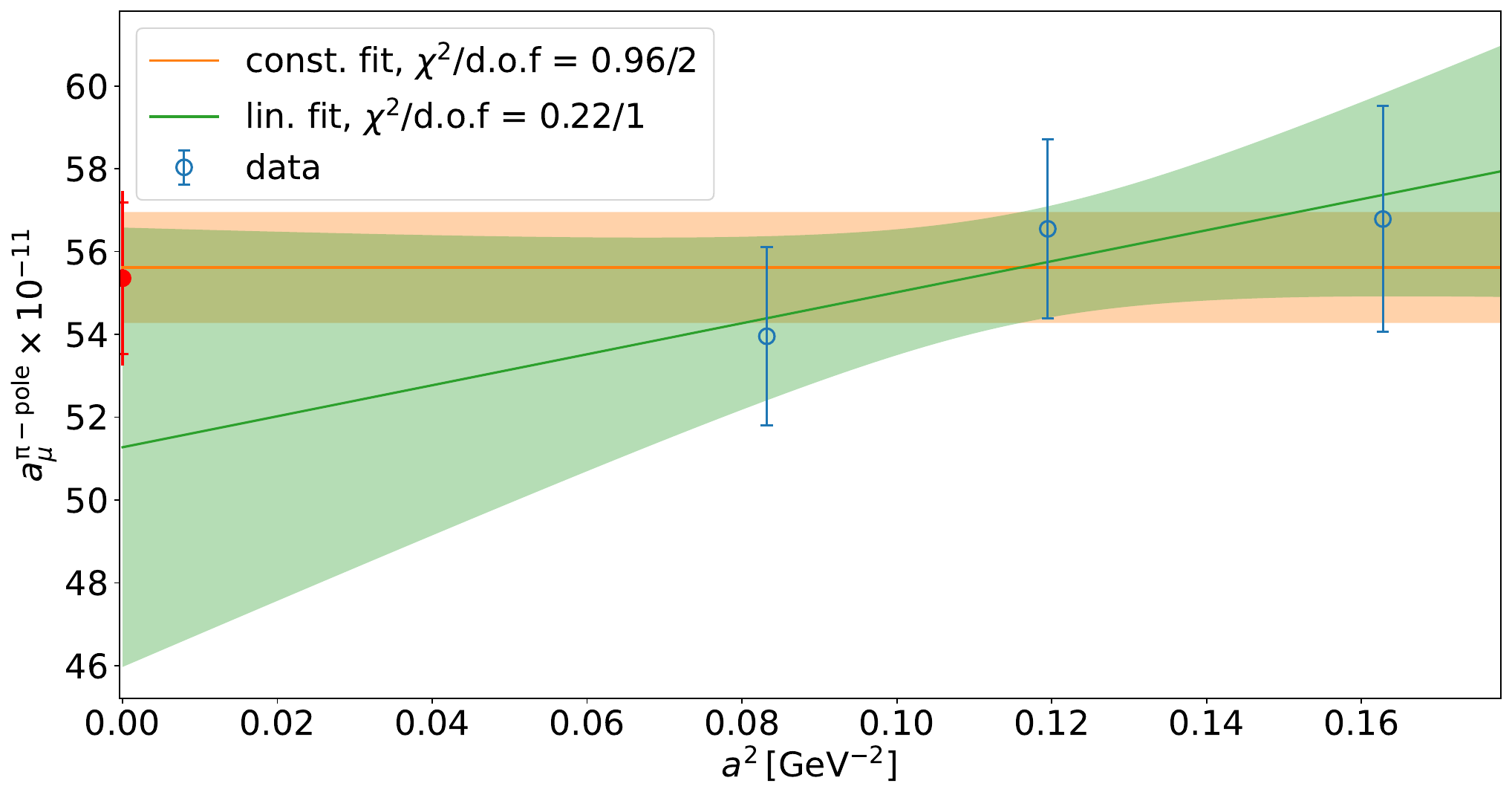}
\caption{Preliminary continuum limit of $\amu{\pi}$ based on both constant and $\mathcal{O}(a^2)$ extrapolations. The extrapolations are averaged using the procedure from \cite{ExtendedTwistedMass:2021gbo}, as discussed in the main text, yielding the continuum estimate $\amu{\pi} = 55.4(1.9)_{\mathrm{fit}}(1.0)_{\mathrm{ctm-syst.}}[2.1]_{\mathrm{tot}}$ indicated by the red marker at $a^2 = 0$.
\label{fig:pion_amu_cont_lim}}
\end{figure}

Data from these three physical-point ensembles allows a continuum
extrapolation to be performed for the value of $\amu{\pi}$. In the
twisted-mass discretization, the leading lattice
artifacts are expected to be
of $\mathcal{O}(a^2)$. For the continuum limit of $\amu{\pi}$ we therefore consider both extrapolation with a simple constant fit versus lattice spacing and a linear fit in $a^2$. As shown in \fig\ref{fig:pion_amu_cont_lim}, both fits are consistent with the lattice data.
To determine a preliminary estimate of the continuum value at the present statistics and set of ensembles, we apply the averaging procedure presented in Ref.~\cite{ExtendedTwistedMass:2021gbo} (see {\eqs}(38)--(43) therein) to combine the constant and linear fits. The resulting estimate of the continuum limit is found to be
\begin{equation}
  \amu{\pi} = 55.4(1.9)_{\mathrm{fit}}(1.0)_{\mathrm{ctm-syst.}}[2.1]_{\mathrm{tot}},
\end{equation}
where the uncertainties are respectively the average uncertainty from the $\tilde A$ and $z$-expansion fits, the additional systematic uncertainty from variation between the constant and linear continuum extrapolations, and the total uncertainty combined in quadrature.

\subsection{Eta meson}
\begin{figure}
\centering
\includegraphics[page=1, width =
1.0\textwidth]{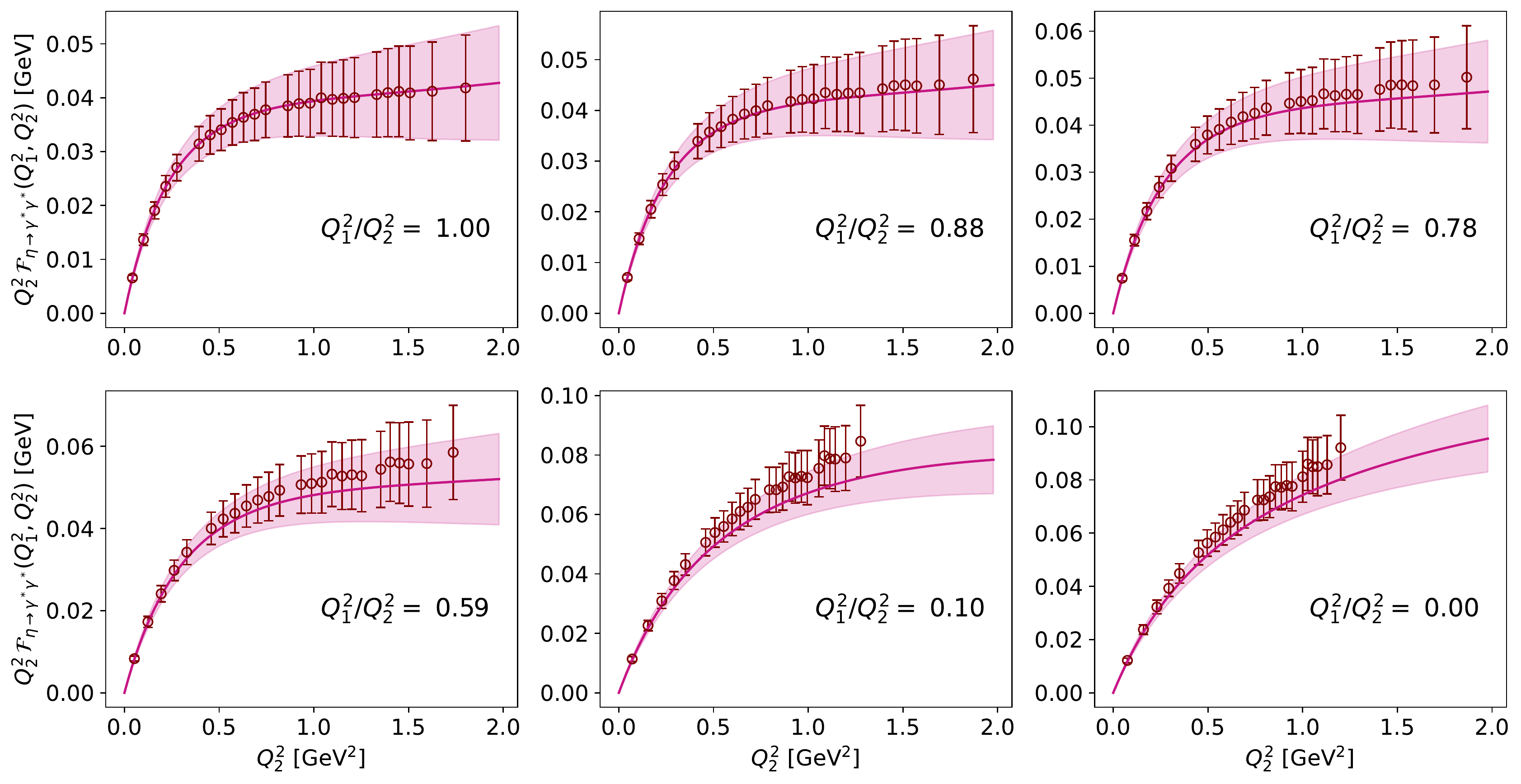}
\caption{Illustration of the $\eta$-meson transition form factor $\vvTFF{\eta}$ and its
  parameterization using the fitted modified $z$-expansion. All data shown is included in the fit. %
\label{fig:zExpansion_eta}}
\end{figure}
As in the case of the pion, for the $\eta$ meson we apply a $z$-expansion fit to extrapolate TFF results to arbitrary kinematics. \fig\ref{fig:zExpansion_eta} shows the example of an $N=2$ $z$-expansion fit on the ensemble cB072.64 using TFF data satisfying $Q_2^2/Q_1^2 \in \{1.0, 0.88, 0.78, 0.59, 0.1, 0.0\}$, where the transition form factors are obtained from a global VMD fit with $\{\tau_\textrm{min}/a, \tau_\textrm{max}/a\} = \{4, 7 \}$ and integration using
$\tau_\textrm{cut}/a$ = 8.
The fit to $\tilde A(\tau)$ is performed with correlation taken into account, yielding $\chi_{\mathrm{VMD}}^2/\textrm{dof} = 0.25$, while the fit to the modified $z$-expansion is uncorrelated with $\chi^2_{z-\mathrm{exp}}/\textrm{dof}=1.03$.
Since $m_\eta \gg m_\pi$, in the case of the $\eta$ meson even the diagonal kinematics include significant contributions from the tails of the integrand $e^{\omega_1 \tau} \tilde A(\tau)$ and thus have significant statistical uncertainties.
\begin{figure}
\centering
\includegraphics[page=2, width =
0.7\textwidth]{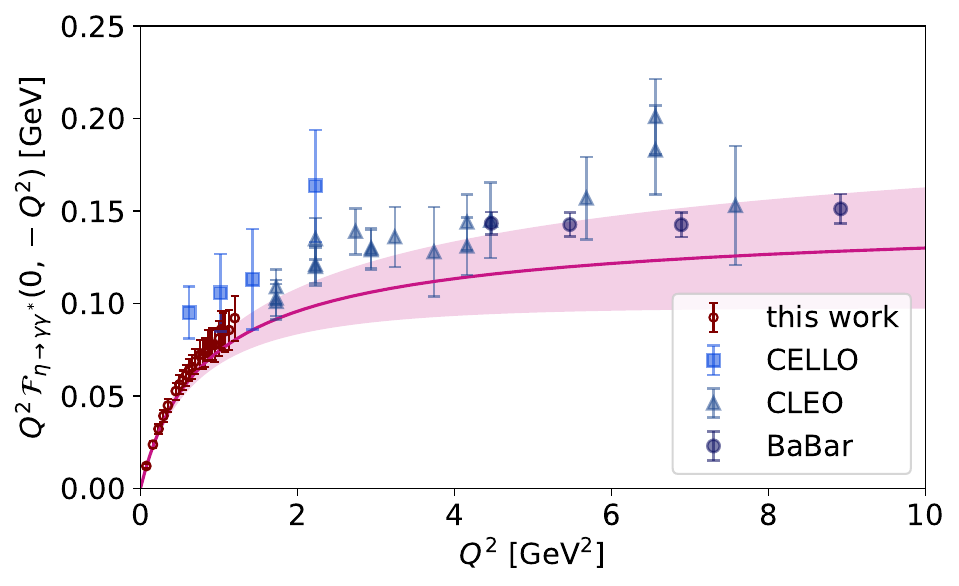}
\caption{Results for the single-virtual $\eta$-meson transition form
  factor $\rvTFF{\eta}$ evaluated using an example set of analysis choices, as described in the main text, versus experimental measurements described in Refs.~\cite{CELLO:1990klc, CLEO:1997fho, BaBar:2009rrj, BaBar:2011nrp}. The red band indicates the result of the $z$-expansion fit to the lattice data shown in red points. As the lattice data and fit are shown for a single choice of analysis parameters, the uncertainties are purely statistical. A full analysis of the form factor itself has now been presented in Ref.~\cite{Alexandrou:2022qyf}. %
\label{fig:zExpansion_eta_exp_comp}}
\end{figure}
In the single-virtual case, we can compare our results for
$\rvTFF{\eta}$, both the data points directly as well as the
$z$-expansion fit, to experimental data from CLEO, CELLO and BaBar,
cf.~\cite{CELLO:1990klc, CLEO:1997fho, BaBar:2009rrj,
  BaBar:2011nrp}. An illustration is shown in
\fig\ref{fig:zExpansion_eta_exp_comp} for the same parameters as in
\fig\ref{fig:zExpansion_eta}. We already find relatively good agreement with the
experimental data despite working at a single lattice spacing.

Using the results of the modified $z$-expansion fits to extend the TFF
$\vvTFF{\eta}(-Q_1^2, -Q_2^2)$ to arbitrary space-like momenta allows us to
evaluate the three-dimensional integral representation given in
\eq\eqref{eq:3drepresentation} to calculate $\amu{\eta}$, as in the pion case.
This is evaluated for each of the $\mathcal{O}(1000)$ combinations of choices of fit range
and fit model in the $\tilde A(\tau)$ fit, the choice of
$\tau_\textrm{cut}$, the choices of samplings of $\vvTFF{\eta}$ in the $(Q_1^2,Q_2^2)$-plane as inputs to the modified $z$-expansion, and the choice of $N \in \{1, 2\}$ for the $z$-expansion. 
Further, for the $\eta$ meson we include all choices of $\tau_{\textrm{cut}}$ satisfying $\tau_{\textrm{cut}} \in [0.16, \, 0.64] \,\mathrm{fm}$.
Using the model averaging procedure of Ref.~\cite{Borsanyi:2020mff}, we obtain the preliminary result
\begin{equation} \label{eq:amu-eta}
\amu{\eta} = 12.7(4.6)_{\mathrm{stat}}(0.7)_{\mathrm{syst}}[4.6]_{\mathrm{tot}} \cdot 10^{-11}
\end{equation}
on the cB072.64 ensemble. We note that the quoted systematic uncertainty in \eq\eqref{eq:amu-eta} only includes the effects of the analysis choices described above, and in particular does not include uncertainties associated with the continuum and infinite-volume limits.

\section{Conclusion and outlook}
\begin{figure}
  \centering
  \includegraphics{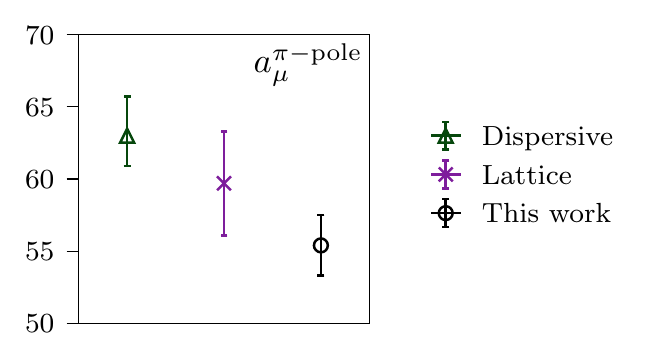} %
  \includegraphics{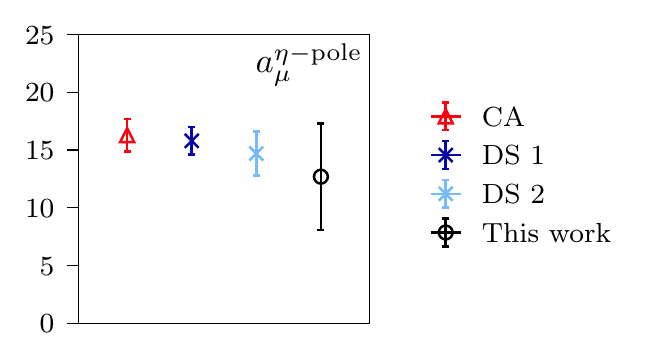}
  \caption{Left: Comparison of the estimate of $\amu{\pi}$ from this work versus an estimate based on dispersion relations~\cite{Aoyama:2020ynm,Hoferichter:2018dmo,Hoferichter:2018kwz} and a prior estimate from lattice QCD~\cite{Gerardin:2019vio}. Right: Comparison of the estimate of $\amu{\eta}$ from this work versus a result derived from Canterbury approximant (CA) fits to experimental data~\cite{Masjuan2017a} and two estimates based on Dyson-Schwinger (DS) equations~\cite{Eichmann2019,Raya2020}. \label{fig:results} }
\end{figure}

Our preliminary result $\amu{\pi} = 55.4(2.1) \cdot 10^{-11}$ may be compared to the recent lattice result $\amu{\pi} = 59.7(3.6) \cdot 10^{-11}$ from Ref.~\cite{Gerardin:2019vio} and the dispersive result $\amu{\pi} = 63.0^{+2.7}_{-2.1} \cdot 10^{-11}$ from Refs.~\cite{Aoyama:2020ynm,Hoferichter:2018dmo,Hoferichter:2018kwz}.
We note that there is a mild tension with the data-driven result, though our analysis is not yet finalized, as discussed below.
Our preliminary result $\amu{\eta} = 12.7(4.6) \cdot 10^{-11}$ on the
cB072.64 ensemble can be compared to the estimate
$\amu{\eta} = 16.3(1.4) \cdot 10^{-11}$ from a Canterbury approximant
fit to experimental data~\cite{Masjuan2017a} and $\amu{\eta} =
15.8(1.2) \cdot 10^{-11}$ and $\amu{\eta} = 14.7(1.9) \cdot 10^{-11}$
using Dyson-Schwinger and Bethe-Salpeter
equations~\cite{Eichmann2019,Raya2020}. Within our uncertainties we find good agreement with these various results already at this single lattice spacing. \fig\ref{fig:results} compares our results for both $\amu{\pi}$ and $\amu{\eta}$ with the aforementioned estimates.
At the moment, the $z$-expansion fits for the pion on cB072.64 are being finalized.
Though this may change the central value of $\amu{\pi}$ in the continuum slightly we do not expect a drastic shift, and also no significant change in the total error. At the same time, the continuum extrapolation strategy is being developed to avoid either a too-conservative estimate with overestimated errors or an overly aggressive estimate dominated by the constant fit.
Finally, future calculations on the cC060.80 and cD054.96 ensembles are planned for the analysis of $\amu{\eta}$, which will allow a continuum estimate of this value directly from ab-initio lattice QCD. An ab-initio value is not yet available for this quantity, so that such a result will provide an important cross-check for data-driven results for $\amu{\eta}$.

\subsection{Acknowledgments}
This work is supported in part by the Sino-German collaborative research center CRC 110 and
the Swiss National Science Foundation (SNSF) through grant No.~200021\_175761, 200021\_208222, and 200020\_200424.
The authors gratefully acknowledge computing time granted on Piz Daint at Centro Svizzero di Calcolo Scientifico (CSCS)
via the projects s849, s982, s1045 and s1133.
The authors also gratefully acknowledge
the Gauss Centre for Supercomputing
e.V. (www.gauss-centre.eu) for funding the project by providing
computing time on the GCS supercomputer JUWELS Booster~\cite{Krause:2019pey} at
the Jülich Supercomputing Centre (JSC).
Part of the results were created within the EA program of JUWELS Booster
also with the help of the JUWELS Booster Project Team (JSC, Atos,
ParTec, NVIDIA).
Ensemble production and measurements for this analysis made use of tmLQCD~\cite{Jansen:2009xp,Deuzeman:2013xaa,Abdel-Rehim:2013wba,Kostrzewa:2022hsv}, DD-$\alpha$AMG~\cite{Alexandrou:2016izb,Alexandrou:2018wiv}, and QUDA~\cite{Clark:2009wm,Babich:2011np,Clark:2016rdz}.
Some figures were produced using \texttt{matplotlib}~\cite{Hunter:2007}.

\bibliographystyle{JHEP}
\bibliography{bibliography}{}

\end{document}